\documentclass[aps,pre,twocolumn,showpacs,amsmath,amssymb]{revtex4}
\makeatletter
\def\@dotsep{4.5}
\makeatother

\usepackage{graphicx}
\usepackage[usenames]{color}

\newlength{\onefig}
\newlength{\twofig}
\newlength{\threefig}
\newcommand{\bq}{\begin{eqnarray}}
\newcommand{\eq}{\end{eqnarray}}
\newcommand{\bqn}{\begin{eqnarray*}}
\newcommand{\eqn}{\end{eqnarray*}}
\newcommand{\beq}{\begin{equation}}
\newcommand{\eeq}{\end{equation}}
     
\newcommand{\der}[2]{\frac{d #1}{d #2}}    
\newcommand{\ca}[1]{\mathcal #1}

\newcommand{\etal}{\textit{et~al.}}

\begin{document}
  
\title{Pressure-energy correlations and thermodynamic scaling in viscous Lennard-Jones liquids}
\author{D. Coslovich}
\email{coslovich@cmt.tuwien.ac.at}
\affiliation{Institut f\"ur Theoretische Physik, Technische Universit\"at
  Wien, Wiedner Hauptstra{\ss}e 8-10, A-1040 Wien, Austria} 
\affiliation{Dipartimento di Fisica Teorica, Universit\`a di Trieste,
  Strada Costiera 11, 34100 Trieste, Italy}
\author{C. M. Roland}
\email{roland@nrl.navy.mil}
\affiliation{Naval Research Laboratory, Code 6120, Washington DC 20375-5342, USA} 
\thanks{Copyright (2009) American Institute of Physics. This article
  may be downloaded for personal use only. Any other use requires
  prior permission of the author and the American Institute of
  Physics.}
\date{\today}
\begin{abstract}
We use molecular dynamics simulation results on viscous binary
Lennard-Jones mixtures to examine the correlation between the
potential energy and the virial. In accord with a recent proposal
[U. R. Pedersen~\etal~Phys. Rev. Lett. {\bf 100}, 015701 (2008)], the
fluctuations in the two quantities are found to be strongly
correlated, exhibiting a proportionality constant, $\Gamma$,
numerically equal to one-third the slope of an inverse power law
approximation to the intermolecular potential function. The
correlation is stronger at higher densities, where interatomic
separations are in the range where the inverse power law approximation
is more accurate. These same liquids conform to thermodynamic scaling
of their dynamics, with the scaling exponent equal to $\Gamma$. Thus,
the properties of strong correlation between energy and pressure and
thermodynamic scaling both reflect the ability of an inverse power law
representation of the potential to capture interesting features of the
dynamics of dense, highly viscous liquids.
\end{abstract}

\maketitle

\section{Introduction}

The myriad phenomena accompanying the supercooling of liquids continue
to intrigue condensed matter scientists. On approaching the glass
transition, the equilibrium, metastable liquid is comprised of fast and
slow moving molecules, which interchange their roles at times on the
order of the structural relaxation. This makes the many-body dynamics
inherently cooperative and thus too complex to be solved by a
first-principles treatment. Various models have been developed to
describe and interpret the dynamic properties, but none rise to the
level of a predictive theory. Nevertheless, progress can be made by
combining experiments that capture some defining characteristic of the
dynamics with computer simulations able to provide a microscopic
interpretation.

One effective simplification in describing the dynamics of
glass-forming liquids is provided by the experimental observation that
structural relaxation times $\tau_\alpha$ and transport coefficients,
encompassing a range of thermodynamic conditions, superpose when
plotted as a function of $\rho^\gamma/T$, where $T$ is temperature,
$\rho$ density, and $\gamma$ a material
constant~\cite{casalini04,albasimionesco04,dreyfus04}. This
thermodynamic scaling of $\tau_\alpha$ has been demonstrated for
dozens of molecular liquids and polymers~\cite{roland05} and extends
with good accuracy from the high temperature Arrhenius region, through
the dynamic crossover, down to the glass
transition~\cite{casalini05a}. The justification for the scaling law
\begin{equation}\label{eqn:scaling}
  \tau_{\alpha} \sim \ca{F}(\rho^\gamma/T)
\end{equation}
draws from early work of Hoover and
coworkers~\cite{hoover71,hoover71a}, who investigated the properties
of fluids having a pair-wise additive intermolecular potential
described by an inverse power law (IPL)
\begin{equation}\label{eqn:ipl}
  u(r) = \epsilon (\sigma/r)^m
\end{equation}
in which $m$ is a constant, $r$ the intermolecular separation, and
$\epsilon$ and $\sigma$ have respective dimensions of energy and
length. Generally, the intermolecular potential for liquids is
represented as a pair-wise additive interaction, with the steep
repulsive part written as an exponential function or an IPL. An
advantage of the IPL is that all reduced thermodynamic and dynamic
properties of such a system can be expressed in terms of the variable
$\rho^{m/3}/T$~\cite{hoover71,hoover71a,hiwatari74}. Simulations of
vitrifying liquids have often employed an IPL
potential~\cite{bernu87,grigera02,demichele04,fernandez07}. Recent
theoretical models of energy landscapes, relevant for description of
the glass transition, are also based on Eq.~\eqref{eqn:ipl} but with
the addition of a density-dependent constant to account for the
long-range attractive
forces~\cite{debenedetti99,shell03,speedy03,shell04}.

At high densities the liquid structure is determined mainly by the
repulsions~\cite{weeks71}, suggesting that if the repulsive potential
were accurately described by an IPL, the local dynamics would depend
only on $\rho^{m/3}/T$, with the empirical scaling exponent $\gamma$
identified with $m/3$. Support for this idea comes from recent
simulations using an $m$-6 Lennard-Jones (LJ) intermolecular potential
\begin{equation}\label{eqn:lj} 
u(r) = 4\epsilon\left[
  {\left( \frac{\sigma}{r} \right)}^{m} - {\left(
    \frac{\sigma}{r} \right)}^6 \right]
\end{equation}
Relaxation times and diffusion constants from these simulations
conform to Eq.~\eqref{eqn:scaling} but with a scaling exponent
$\gamma$ that is not equal to
$m/3$~\cite{budzien04,tsolou06,coslovich08}. For polymers this is due
in some measure to the effect of the intramolecular part of the
potential~\cite{roland06}. More generally, a systematic study of LJ
particles~\cite{coslovich08} showed that $\gamma$ follows closely the
steepness of the repulsive core. In the latter system, however, the
scaling exponent $\gamma$ is always larger than $m/3$, due to
contributions of the attractive term in
Eq.~\eqref{eqn:lj}~\cite{ben-amotz03,bailey08a,bailey08b}. $\gamma$ is
equal to the slope of an IPL fitted over the typical distances of
closest approach between particles probed in the highly viscous
regime. These interatomic separations are defined by values of $r$
between the first nonzero value of the radial distribution function
and the position of the half-height of the first
peak~\cite{coslovich08}.

In addition to giving rise to the scaling of the dynamics
[Eq.~\eqref{eqn:scaling}], adequacy of the IPL approximation for
highly viscous liquids suggests the existence of a correlation between
equilibrium fluctuations of the configurational parts of energy and
pressure; i.e., between the potential energy $U$ and the virial
$W$. This expectation follows from the exact correlation between
fluctuations in $U$ and $W$ for an IPL~\cite{bailey08a}. Recently,
Dyre and coworkers showed from simulations that the potential energy
and the virial strongly correlate for various non-associated
materials, with correlation coefficients exceeding
0.9~\cite{pedersen08,bailey08a}. Liquids exhibiting both thermodynamic
scaling and correlation between $U$ and $W$ may exhibit other
interesting properties, such as sufficiency of a single parameter to
describe their temperature-dependent
viscoelasticity~\cite{ellegaard07,bailey08c}. In this work we assess
the correlation of $U$ and $W$ in viscous LJ liquids having different
repulsive interactions; i.e., different $m$ in Eq.~\eqref{eqn:lj}.
These are the same LJ liquids previously shown to conform to
thermodynamic scaling of their diffusion
coefficients~\cite{coslovich08}. Thus, from our results we can test
the conjecture of Pedersen~\etal~\cite{pedersen08} that liquids whose
dynamics follow the thermodynamic scaling are strongly correlating and
\textit{vice versa}.

\section{Models}\label{sec:models}

The models considered in this work consist of additive, equimolar
binary mixtures of 500 particles interacting through the $m$-6 LJ
potential
\begin{equation}\label{eqn:ljm} 
u_{\alpha\beta}(r) = 4\epsilon_{\alpha\beta}\left[
  {\left( \frac{\sigma_{\alpha\beta}}{r} \right)}^{m} - {\left(
    \frac{\sigma_{\alpha\beta}}{r} \right)}^6 \right]
\end{equation}
where $\alpha, \beta = 1,2$ are indexes of species. The mixtures
studied have a size ratio $\sigma_{11}/\sigma_{22} = 0.64$, masses
$m_1/m_2 = 1.0$, and a unique energy scale
$\epsilon_{12}/\epsilon_{11} = \epsilon_{22}/\epsilon_{11} = 1.0$. We
refer to them herein as AMLJ. As in previous work~\cite{coslovich08},
we consider different repulsive interactions: $m=8$, 12, 24, and
36. We also study two other LJ liquids having $m=12$: one proposed by
Wahnstr\"om~\cite{wahnstrom}, which we denote as WAHN, and the
non-additive mixture of Kob and Andersen~\cite{ka1}. This latter,
called herein BMLJ, is among the most widely used in molecular
dynamics simulations of the glass transition. In the following reduced
LJ units will be used, assuming $\sigma_{11}$, $\epsilon_{11}$ and
$\sqrt{m_1\sigma_{11}^2/\epsilon_{11}}$ as units of distance, energy
and time respectively. All samples are cooled isobarically at
pressures $P=5$, 10, and 20 using Berendsen thermostat and barostat
during equilibration. The production runs are then performed in the
$NVE$ ensemble using the Velocity-Verlet algorithm. Further details on
the simulations of AMLJ models can be found in
Refs.~\cite{coslovich08,coslovich07a,coslovich07b}.

We remark that the actual pressure varies with $m$ even though the
numerical values of $P$ are the same, because the depth of the
potential increases with $m$. A more appropriate set of reduction
parameters for the potential in Eq.~\eqref{eqn:ljm} is given by the
position $r^*$ and the width $u^*=u_{11}(r^*)$ of the minima of
$u_{11}(r)$. In terms of these parameters, the effective pressure is
$P^*=P(u^*/{\sigma^*}^3)$, where $P$ is expressed in reduced LJ
units. Considering the variation of $u^*$ and $\sigma^*$ for different
$m$, the pressure $P$ for $m=36$, for example, should be increased by
about a factor of 2.5 to match that of $m=12$.

\section{Results}

\begin{figure}[tb]
\begin{center}
\includegraphics*[width=\twofig]{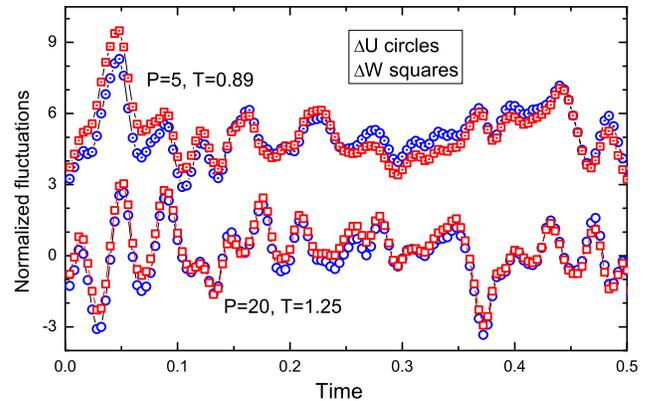}
\end{center}
\caption{\label{fig:uw} Normalized instantaneous fluctuations of the
  potential energy $(U-\langle U\rangle)/\sqrt{\langle\Delta
    U^2\rangle}$ (circles) and virial $(W-\langle
  W\rangle)/\sqrt{\langle\Delta W^2\rangle}$ (squares) for the AMLJ
  model with $m = 36$ at the indicated pressure and temperature. The
  ordinate of the upper curves has been shifted 5 units for clarity.}
\end{figure}

\begin{figure*}[tb]
\begin{center}
\includegraphics*[width=\onefig]{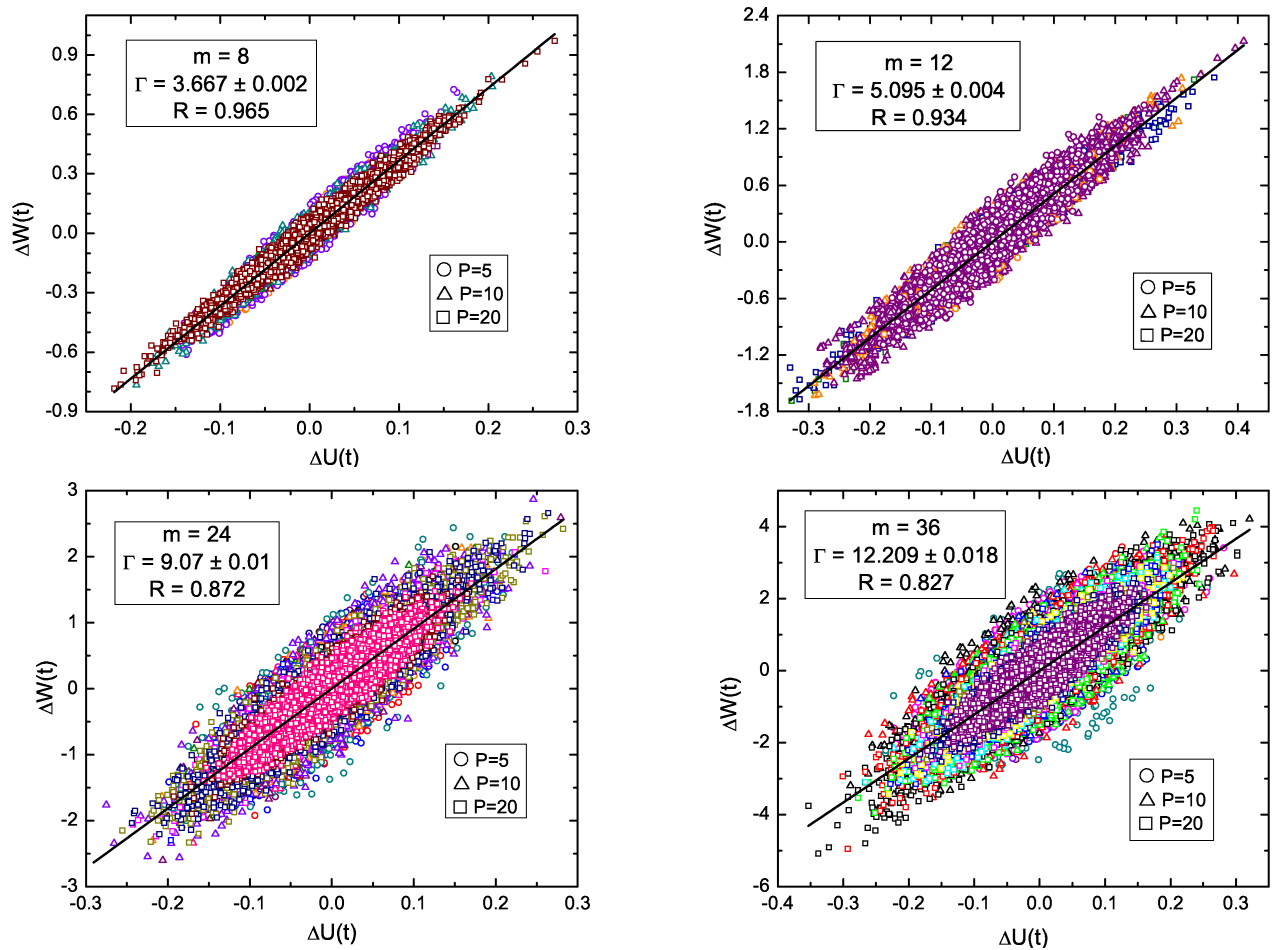}
\end{center}
\caption{\label{fig:m} Fluctuations of the virial versus
  those of the potential energy in the AMLJ model with $m=8$, 12, 24,
  and 36. For each $m$, results for all pressures $P=5$, 10, and 20
  and all temperatures studied are shown. Colors in the
    online version indicate datasets corresponding to different state
    points. The corresponding range of densities is $1.04 \leq \rho
  \leq 2.06$ for $m=8$, $0.87 \leq \rho \leq 1.78$ for $m=12$, $0.93
  \leq \rho \leq 1.70$ for $m=24$, and $0.77 \leq \rho \leq 1.71$ for
  $m=36$. The solid lines represents the least squares linear fits
  with the correlation coefficient $R$ indicated.}
\end{figure*}

Figure~\ref{fig:uw} is a plot of the instantaneous values of the
normalized fluctuations in the virial and the potential energy for the
AMLJ liquid with $m=36$ for two selected conditions of $T$ and $P$. As
discussed below, this material exhibits the weakest correlation of $U$
and $W$ within the class of systems and range of state conditions
investigated herein; nevertheless, a relationship between the two
quantities is evident. The degree of correlation can be assessed by a
variety of means. We calculate the Pearson correlation coefficient
\begin{equation}\label{eqn:R}
R=\frac{\langle\Delta U\Delta W\rangle}
{\sqrt{\langle(\Delta U)^2\rangle\langle(\Delta W)^2\rangle}}
\end{equation}
from linear regression of $\Delta U=U-\langle U\rangle$ and $\Delta
W=W-\langle W\rangle$, with the large number of data points, typically
$2\times 10^5$ per sample, enhancing statistical reliability. The
determination of $R$ is carried out simultaneously for all state points
over which thermodynamic scaling was observed in
Ref.~\cite{coslovich08}; that is, for each pressure ($P = 5$, 10, and
20) at temperatures corresponding to normal liquid conditions down to
the slow-dynamics regime.

\begin{figure}[tb]
\begin{center}
\includegraphics*[width=\twofig]{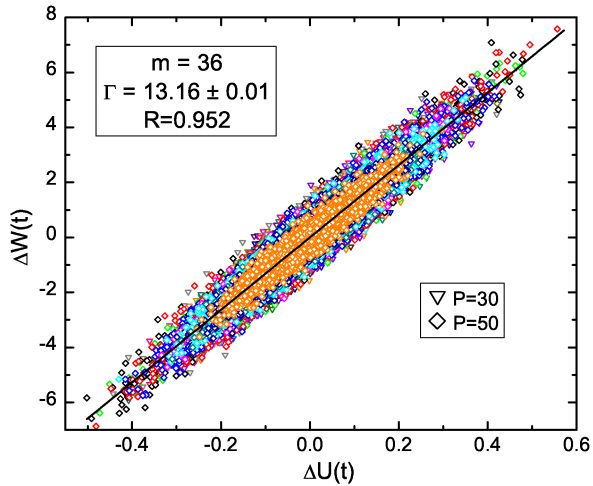}
\end{center}
\caption{\label{fig:m36P} Fluctuations of the virial versus
  those of the potential energy for the AMLJ model with $m=36$ at
  pressures $P=30$ and 50 and all $T$ studied (yielding $1.20 \leq
  \rho \leq 1.75$). These correspond to pressures $P^*$ (see
  Section~\ref{sec:models}) comparable to the values for $m=12$ at
  $P=10$ and 20.}
\end{figure}

In Fig.~\ref{fig:m} the fluctuations in the virial are
plotted versus those in the potential energy for AMLJ liquids with $m
= 8$, 12, 24, and 36. The obtained $R$ are listed in
Table~\ref{table}, from which two observations can be made: The
correlation coefficients are all close to unity, indicating
substantial correlation. This concurs with the results of
Bailey~\etal~\cite{bailey08a} for simulations of one-component LJ
liquids with $m = 12$ and of BMLJ particles. Second, the magnitude of
$R$ decreases from \textit{ca}. 0.97 to 0.83 with increasing magnitude
of the repulsive exponent. This ostensibly suggests that liquids with
steeper intermolecular repulsions are less strongly
correlating. However, as discussed in Section~\ref{sec:models}, the
actual pressure regime explored by the particles in the simulation
changes with $m$. Thus, larger $m$ corresponds to smaller effective
$P$ and $\rho$. Since the correlation improves at higher
density~\cite{bailey08a}, reflecting the greater accuracy of the IPL
approximation, the smaller $R$ for larger $m$ can be ascribed to the
differences in thermodynamic conditions. We demonstrate this in
Fig.~\ref{fig:m36P}, showing results for $m=36$ calculated for $P=30$
and 50. At these higher pressures the correlation coefficient is
similar to the value of $R$ obtained for $m=12$ at $P=10$ and 20. This
confirms that the poorer correlation (smaller $R$) at higher $m$ is
due to the lower effective pressure, and thus larger mean nearest
neighbor distances, rather than to the steepness of repulsion
\textit{per se}.

\begin{figure}[tb]
\begin{center}
\includegraphics*[width=\twofig]{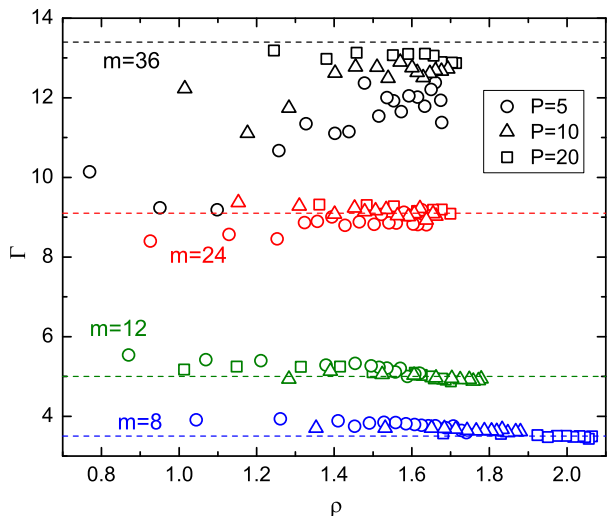}
\end{center}
\caption{\label{fig:gamma} Least squares slope $\Gamma$ calculated
  along individual isobars shown as a function of density
  $\rho=\rho(T,P)$. Results are shown for the AMLJ model with $m=8$
  ($0.18\leq T\leq 2.0$), 12 ($0.41\leq T\leq 3.0$), 24 ($0.75\leq
  T\leq 4.0$), and 36 ($0.87\leq T\leq 6.0$). The horizontal dashed
  lines denote the corresponding values of the scaling exponent
  $\gamma$ (from Ref.~\cite{coslovich08}).}
\end{figure}

For an exact IPL $\der{W}{U} = m/3$~\cite{bailey08a} and regression of
$W(U)$ yields the value of $m/3$ as the slope of the fitted line. More
generally, while the fact that $R \sim 1$ affirms correlation between
$U$ and $W$, this is not a proof of proportionality of the two
quantities~\cite{anscombe73}. However, it can be observed that the
scatter in the plots of Fig.~\ref{fig:m} is normally distributed
(random scatter without systematic trends), thereby justifying an
interpretation of the slope, $\Gamma$, of the fitted lines as a
measure of $m/3$. Results for all simulations, including the BMLJ and
WAHN, are listed in Table~\ref{table}, where it can be seen that there
is good correspondence between the slope $\Gamma$ and the scaling
exponent $\gamma$; that is, strongly correlating liquids conform to
thermodynamic scaling, thus confirming the results of
Ref.~\cite{schroeder08}.

We can examine the relationship between pressure and energy
fluctuations in more detail by evaluating the correlation for each
state point individually. Figure~\ref{fig:gamma} shows $\Gamma(T,P)$,
obtained from linear regression, as a function of
$\rho=\rho(T,P)$. Results for AMLJ systems with $m=8$, 12, 24, and 36
are shown in the figure, along with the scaling exponents (indicated
by the dashed line) obtained from superpositioning of the diffusion
constants for these liquids. Interestingly, at fixed $m$,
$\Gamma(T,P)$ essentially collapse onto a single horizontal curve when
plotted versus $\rho$, with only some variation at lower $T$ and
$P$. These changes in $\Gamma(T,P)$ reflect the fact that the IPL
approximation depends weakly on the state point, with the fluctuations
of $\Gamma(T,P)$ for different state points mostly dictated by density
variations (again excepting $m=36$ for the reasons discussed
above). For each $m$, the mean of $\Gamma(T,P)$ over all state
conditions is equivalent within the error to the $\Gamma$ obtained
from a global fit of $\Delta W$ versus $\Delta U$ (see
Table~\ref{table}).

\begin{table}[tb]
\caption{\label{table} Summary of exponents from $U$-$W$ correlations
  ($\Gamma$) and from thermodynamic scaling of diffusion ($\gamma$)
  for various LJ liquids. The correlation coefficient $R$ obtained
  from linear regression of $W(U)$ data is also included. Statistical
  uncertainties on $\Gamma$ correspond to one standard deviation in
  the ``mean'' case, and to the error associated to linear regression
  in the ``global'' case.}
\begin{ruledtabular}
\begin{tabular}{lccccc}
 & $m/3$ & $\gamma$ & $\Gamma$ (global) & $R$ & $\Gamma$ (mean) \\
\hline
AMLJ36 & 12  & 13.4 $\pm$ 0.2 & 12.21  $\pm$ 0.02 &   0.827 & 12.1 $\pm$ 0.9 \\
AMLJ24 & 8   & 9.1  $\pm$ 0.1 &  9.07  $\pm$ 0.01 &   0.872 &  9.0 $\pm$ 0.2 \\
AMLJ12 & 4   & 5.0  $\pm$ 0.1 &  5.095 $\pm$ 0.004 &   0.934 &  5.10 $\pm$ 0.17 \\
AMLJ8  & 2.7 & 3.5  $\pm$ 0.1 &  3.667 $\pm$ 0.002 &   0.965 &  3.67 $\pm$ 0.13 \\
BMLJ   & 4   & 5.0  $\pm$ 0.1 &  5.087 $\pm$ 0.003 & 0.943 &  5.10 $\pm$ 0.15 \\
WAHN   & 4   & 5.0  $\pm$ 0.1 &  5.052 $\pm$ 0.003 & 0.978 &  5.16 $\pm$ 0.19 \\
\end{tabular}
\end{ruledtabular}
\end{table}

One final comment concerns the link between the $\Gamma$ from the
$U$-$W$ correlations and the thermodynamic scaling exponent $\gamma$.
In Ref.~\cite{coslovich08} we found that a single value of $\gamma$
gave excellent superpositioning of the dynamic data. This means that
any change in ``local'' $\gamma$ with $T$ and $P$ must be small (this
issue was examined quantitatively for simulated liquid silica data in
Ref.~\cite{legrand07}). Even for the more poorly correlating liquid
($m=36$), the estimated uncertainty in $\gamma$ is only about 15\%
($\pm 0.2$). On the other hand, the slopes describing $U$-$W$
correlations display a somewhat wider variation upon changing state
parameters, especially at low density and temperature. This may be
related to the sensitivity of pressure-energy fluctuations to the
shape of the pair potential for distances $r$ around and beyond the
first peak in the radial distribution function (e.g., departures from
the IPL form)~\cite{bailey08b}. Further investigations on the role of
the attractive tail of the potential on $U$-$W$ correlations may be
required to clarify this point.

\section{Conclusions}

In summary, we have shown that for viscous $m$-6 LJ liquids conforming
to thermodynamic scaling of their dynamics, there is a strong
correlation between the virial and the potential energy. This property
is maintained for systems ranging from relatively soft particles ($m =
8$) to those approaching the hard sphere limit ($m = 36$). The
correlation deteriorates for lower densities, as the range of the
fluctuations extends to large $r$, for which the IPL approximation
breaks down. The slopes obtained from linear regression of the virial
vs. potential energy data are in good agreement with the scaling
exponents yielding superpositioning of dynamic data, supporting the
conjecture that pressure-energy correlations and thermodynamic scaling
have a common origin in the IPL approximation of the interaction
potential. Our results are in accord with the recent work of Dyre and
coworkers~\cite{pedersen08,bailey08a,bailey08b,schroeder08} on LJ
particles and suggest the utility of the IPL approximation in
describing essential features of the dynamics of dense, highly viscous
liquids.

\begin{acknowledgements} 
Computational resources were obtained through a grant within the
agreement between the University of Trieste and CINECA
(Italy). D.~C. acknowledges financial support by the Austrian Science
Fund (FWF) (Project number: P19890-N16). The work at NRL was supported
by the Office of Naval Research.
\end{acknowledgements}


\end{document}